\documentclass[10pt,twoside]{article}
\usepackage{Latex-document}
\almvol{00}
\almttone{Frontiers of Science Awards}
\almtttwo{for Math/TCIS/Phys}
\firstpage{1}
\usepackage[english]{babel}
\usepackage[utf8]{inputenc}

\usepackage{multicol}
\usepackage{graphicx}
\usepackage{bm}
\usepackage{tabu}

\usepackage{amsmath}
\usepackage{amssymb}

\numberwithin{equation}{section}

\begin{document}

\markboth{\hfill{\rm Bernd Rosenow and Bertrand I. Halperin} \hfill}{\hfill {\rm  Braids and Beams: Exploring Fractional Statistics with  Anyon Colliders  \hfill}}

\title{Braids and Beams: Exploring Fractional Statistics with Mesoscopic Anyon Colliders}

\author{Bernd Rosenow and Bertrand I. Halperin}

\begin{abstract}
Anyon colliders -- quantum Hall devices where dilute quasiparticle beams collide at a quantum point contact -- provide an interferometer-free probe of  anyonic exchange phases through current cross correlations.
Within a non-equilibrium bosonization framework, the normalized cross-correlations take a universal form depending only on the exchange phase and the dynamical exponent, enabling experimental demonstration of anyonic statistics. 
This result can be interpreted as time-domain interference -- braiding in time rather than spatial exclusion or real-space interferometry.
Extension to hierarchical states shows that the semiclassical step-function description of quasiparticles fails at large statistical angles.
Introducing a finite soliton width resolves this issue and enables quantitative modeling of charge-$e/5$ quasiparticle collisions. 

\end{abstract}

\maketitle

\setcounter{tocdepth}{1}
\tableofcontents

\section{Introduction}

The statistics of quantum mechanical particles is described by the symmetry of the many-particle wave function under the exchange of two particle coordinates. Whereas the exchange gives rise to a plus sign for bosons, there is a minus sign
for fermions. This sign difference leads to fundamentally distinct many-body behavior: fermions form Fermi liquids and metals, while bosons may undergo Bose-Einstein condensation into a superfluid.
In two-dimensional systems, the notion of quantum statistics can be extended in interesting ways \cite{LeMy77,Laughlin83,Halperin84,ArScWi84}. 


As an example, consider the fractional quantum Hall (FQH) effect. 
A low energy state may contain a small number of fractionally charged quasiparticles (qps) on top of the ideal ground state.  Its time evolution may then be described by an effective wavefunction $\Psi_{\rm{eff}}$ that depends on the positions of the qps and an effective Hamiltonian which acts on it. For Abelian FQH states, counterclockwise interchange of two qps, around a path that doe not enclose any other qps, will
multiply $\Psi_{\rm{eff}}$ by a complex phase  $e^{i \theta_a}$, where $\theta_a $ denotes the statistical angle.
 Demonstrating this fractional statistics experimentally, and thereby establishing the anyonic nature of FQH qps, has long been a central challenge  that has driven numerous theoretical proposals and experimental efforts \cite{Feldman22}.

%
\begin{figure}[t]
\includegraphics[width=0.9\linewidth]{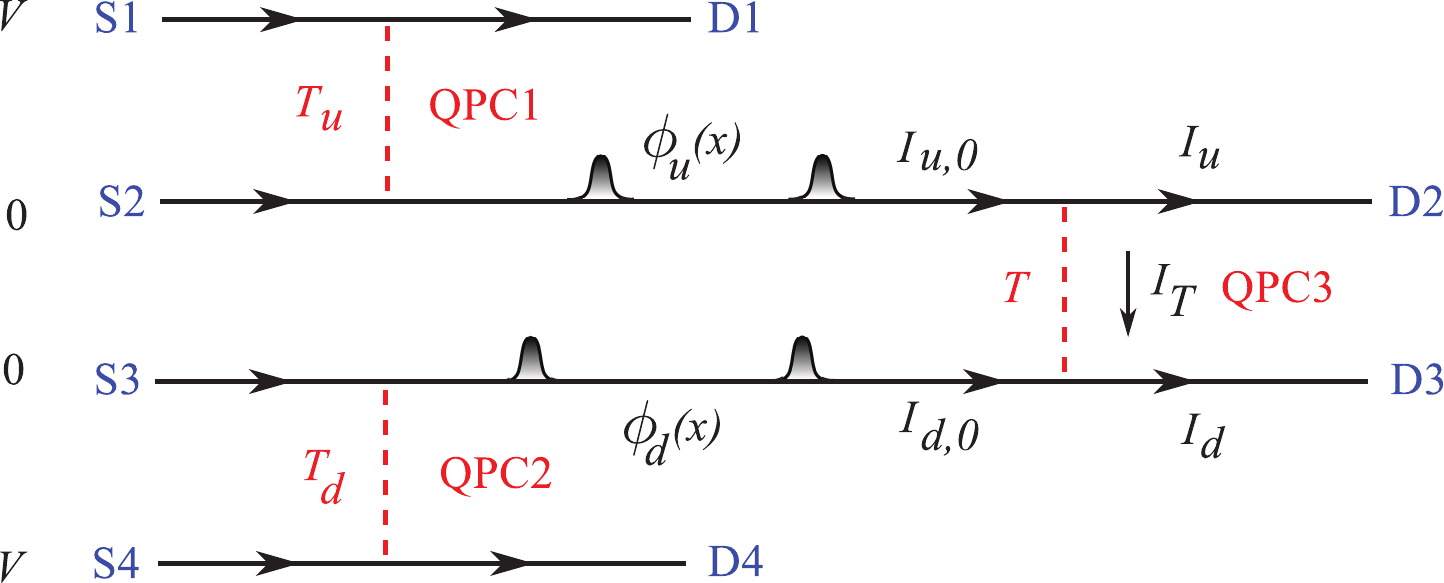}
\caption{  Schematic setup of an anyon collider. Dilute anyon beams are generated at QPC1 and QPC2 with small transmission probabilities $T_u$ and $T_d$, collide at QPC3, and are collected at drains D2 and D3. The resulting current cross correlations are measured at drains D2 and D3 (adapted from Ref.~\cite{Ro+16}.}
\label{fig:setup}
\end{figure}
%

Collision experiments with anyons offer a powerful route to probe their fractional statistics
  \cite{Smitha03,ViCo10,Campagnano+12,Campagnano+13}, conceptually based on the  pioneering work of Hanbury Brown and Twiss (HBT) \cite{Hanbury56}. Numerous theoretical works have proposed extracting signatures of fractional statistics from current correlations
  \cite{Sa+01,Kim05,Martin05,Kim06,Ardonne08,Safi20}. Further progress was made with the proposal of an anyon collider  \cite{Ro+16}, where two  quantum point contacts generate  dilute beams of  anyons, which  collide at a third quantum point contact (see Fig.~\ref{fig:setup}).  
The signature of  anyonic statistics is contained in the cross correlations of  currents defined in drain contacts after the collision QPC. Such a setup was  realized experimentally in
\cite{Bartolomei.2020,Lee.2023,Ruelle.2023,Glidic.2023}, providing evidence for anyonic statistics.
Crucially, the observed signal reflects time-domain interference of anyons and does not require a conventional interferometer
\cite{Han+16,Lee+20,Lee+22,Schiller23}.  Further experimental variants have been realized: quasiparticle Andreev scattering at $\nu=1/3$~\cite{Glidic2023b}, and, at integer filling, time-domain braiding between fractional pulses and particle-hole excitations~\cite{Glidic2024}. 
The dynamics of time-domain braiding have been investigated experimentally in \cite{Ruelle+25}, and theoretical proposals have suggested extracting the braiding phase from  signatures in effective transport coefficients~\cite{zhang25a,zhang25b}.


To build some intuition for signatures in current cross-correlations in a mesoscopic collider, 
 it is instructive to first consider free fermions,  which can be described in a scattering matrix formalism. 
 We denote the tunneling probability at the collision QPC by $T$. 
 Assuming $eV>0$ in Fig.~\ref{fig:setup}),
 the incoming fermions are described by a double-step distribution $f_\alpha(\epsilon) = \theta(-\epsilon) + T_\alpha \, \theta(\epsilon) \theta(eV - \epsilon)$  arising from injection at the source QPCs, where $\alpha = u, d$.
We then obtain that the zero-frequency cross-correlations
are \cite{Ro+16}
%
\begin{equation}
 \langle \delta I_u \delta I_d\rangle_{\omega=0} = - T (1-T)eV(e^2/h)(T_u - T_d)^2 \  . 
\end{equation}
%
The cross-correlations vanish for equal bias currents  with $T_u =  T_d$,  reflecting the absence of cross-correlations in the case  of fermionic statistics.

To connect this result to the quantum statistical properties of fermions, 
we consider a classical lattice model with two chains, which has occupation probabilities $T_u$ and $T_d$ on a given lattice
site for incoming particles. In each time step, particles move one lattice
site forward, and we model the collision QPC by a special lattice site, where single particles can tunnel between chains with a probability $T$. 
When two particles arrive simultaneously at the collision QPC, 
the probability for tunneling is reduced to $(1-p)T$. Due to the Pauli exclusion principle, fermions correspond to $p=1$, and for a general value 
of $p$ we find current cross-correlations \cite{Ro+16}
%
\begin{equation}
\langle \delta I_u \delta I_d\rangle_{{\rm classical}, \omega=0} \ = \ - T V{e^3 \over h}(T_u - T_d)^2 - 2 T V{e^3 \over h}(1-p)T_u T_d \ .
\label{classical_correlation.eq}
\end{equation}
%
For two incoming beams with equal occupation probabilities $T_u=T_d$, the
cross correlations are proportional to $-(1-p)$ and thus carry information
about the two-particle exclusion statistics. 
The result is proportional to $T_uT_d$, since collisions occur only when two particles reach the QPC simultaneously.
 As we will see, the quantum-mechanical result differs markedly and cannot be explained by classical exclusion statistics alone.

\section{Anyon Collider}

We consider first the simplest case of a Laughlin state at filling $\nu = 1/m$, which was the main focus of Ref \cite{Ro+16}. Here we have 
$e^*=e/m, \, \theta_a = \pi/m$. Tunneling of qps between the two edges at a QPC maybe characterized by a tunneling operator $A$ and associated tunneling current operator $I_T$, which we may write in terms of boson fields as
%
\begin{equation}
A(t) \ = \ \zeta e^{i \phi_u(0,t) - i \phi_d(0,t)} \ \ , \ \ \ I_T \ = \ ie^\star \left( A^\dagger - A\right)  \ , 
\label{Eq3} 
\end{equation}
%
with tunneling amplitude $\zeta$.
The charge density on edge $\alpha= (u,d)$ is  $\rho_{\alpha} = \partial_x \phi_{\alpha}/2 \pi$ \cite{Giamarchi_book}, and the boson fields satisfy the equal-time commutation relation $[\phi_\alpha(x), \phi_\beta(y)] = i (e^\star/e) \pi \delta_{\alpha, \beta} {\rm sign}(x-y)$. 
To leading order in $|\zeta|^2$, the average tunneling current and its zero-frequency fluctuations are given by
%
\begin{subequations}
\begin{eqnarray}
\hspace*{-0.5cm} \langle I_T \rangle & = & e^\star \int_{-\infty}^\infty \! \! \!   dt \langle [A^\dagger(0), A(t)]\rangle_0 \ ,  \\[.5cm]
\hspace*{-0.5cm}\langle (\delta I_T)^2\rangle_{\omega=0} &= & (e^\star)^2 \int_{-\infty}^\infty  \! \! \! dt \, \langle\{ A^\dagger(0) , A(t)\}\rangle_0  \  ,
\end{eqnarray}
\end{subequations}
%
  with $[\,\cdot\,,\,\cdot\,]$ the commutator and $\{\,\cdot\,,\,\cdot\,\}$ the
anticommutator,
and $\langle \, \cdot \, \rangle _0$ is a non-equilibrium expectation value in the absence of tunneling at the collision QPC. 

In non-equilibrium settings, averages of vertex operators do not reduce to simple exponentials of bosonic two-point functions. 
 Within a semiclassical approach (described below) the quantum mechanical
displacement fields and the classical counting fields are statistically
independent, and the correlator factorizes into an equilibrium power-law contribution and a separate non-equilibrium factor, 
 %
\begin{equation}
\langle A(t) A^\dagger(0)\rangle_0 \ = \  e^{i \pi \delta\,  {\rm sign}(t)} \, {\tau_c^{2 \delta} \over 
|t|^{2 \delta}}  \times  \langle A(t) A^\dagger(0)\rangle_{\rm non-eq} \ ,
\label{quantumcorrelation.eq}
\end{equation}
%
 where $\tau_c$ is a short time cutoff. The dynamical exponent $\delta$ is equal to $e^*/e = 1/m$ for the simple edge mode, but could differ in edge structure with additional 
 counter-propagating modes
coupled to the charge mode  \cite{RoHa02}.  

Evaluating the non-equilibrium contribution requires a non-perturbative treatment of tunneling at QPC1 and QPC2, since edge correlation functions decay only slowly. 
We employ non-equilibrium bosonization, adapted to anyonic quasiparticles \cite{Levkivskyi16,Ro+16}, which treats tunneling at the injection QPCs non-perturbatively and captures the full power-law behavior of the correlation functions.
 This approach is directly
applicable to steady states driven far from equilibrium and will be the basis of
our analysis.

The characteristic temporal width of an injected anyon wave packet is $\tau_s = \hbar/(e^\star V)$.  
  We consider the dilute-beam regime, in which $\tau_s$ is much smaller
than the typical time between tunneling events, $e^\star/I_0$. 
In this limit we decompose the boson field as $\phi_{\alpha} = \phi_{\alpha}^{(0)} + 2 \pi \lambda N_\alpha$, where $\lambda=1/m=e^*/e= \theta_a / \pi$ at $\nu=1/m$.
The equilibrium part produces the power law in Eq.~\eqref{quantumcorrelation.eq}, while the non-equilibrium part is represented semiclassically by a counting field $N_\alpha(t)$ that increases by a step of height 
$\lambda$ whenever a quasiparticle passes the collision QPC on edge $\alpha$.


In the dilute (Poissonian) limit, tunneling events  occur statistically independently. 
The non-equilibrium factor is then given by the generating function of the independent Poisson processes on the two edges \cite{Levkivskyi16,Ro+16}, and one finds
 %
\begin{eqnarray}
\langle A(t) A^\dagger(0)\rangle_{\rm non-eq} & = & |\zeta|^2    \exp\!\left[- {\langle I_{u,0}\rangle \over e^\star}\! \left(1\! - \!  e^{- 2  i \theta_a}\right)\! t \right]  \exp\!\left[
- {\langle I_{d,0}\rangle \over e^\star}\! \left(1\! -\!  e^{2  i \theta_a}\right)\! t\right]  \ \ \ \ \ \ \ \ 
\end{eqnarray}
%
  with $\langle I_{\alpha,0}\rangle/e^\star = \langle\partial_t N_\alpha\rangle$.
Using Eq.~\eqref{quantumcorrelation.eq} (above), one obtains scaling forms for
the tunneling current and its fluctuations \cite{Ro+16},
   %
\begin{eqnarray}
\langle I_T\rangle 
& = &\! C \sin(\pi \delta) {\rm Im}\left(I_++ {iI_-\over \tan \theta_a}\right)^{2\delta-1} \! \!   \left[1 + O(\tau_c)\right] ,
\end{eqnarray}
%
 and  
%
\begin{eqnarray}
{\langle \delta I_T^2\rangle_{\omega=0}\over  (e^\star)^2}
 & =&  { C \over e^\star} \cos(\pi \delta) {\rm Re}\left(I_++ {iI_-\over\tan \theta_a}\right)^{2\delta-1}\! \! \!  \left[1 + O(\tau_c)\right] \ ,
  \label{tunnelnoise.eq}
\end{eqnarray}
%
where  $O(\tau_c) \to 0 $ for $\tau_c \to 0$, 
$C=e^\star 4 |\zeta|^2 \tau_c^{2 \delta} [\pi(1 - \cos 2 \theta_a)/e^\star]^{-1 + 2 \delta} \Gamma(1 - 2 \delta)$, 
$I_+ = |\langle I_{u,0}\rangle|   + |\langle I_{d,0}\rangle| $, and $ I_- = \langle I_{u,0}\rangle   - \langle I_{d,0}\rangle$. 
 If one were to use these formulas with a parameter 
  $\theta_a > \pi/2$ the effective drive in $\langle I_T\rangle$ would change sign, and the current may even appear negative, signaling a breakdown of the step-function description. Such a situation can occur in states heirarchical states, with multiple edge modes, where it is resolved by incorporating a finite soliton width.

\begin{figure}[t!]
    \centering
    \begin{minipage}[c]{0.48\linewidth}
    \includegraphics[width= \linewidth]{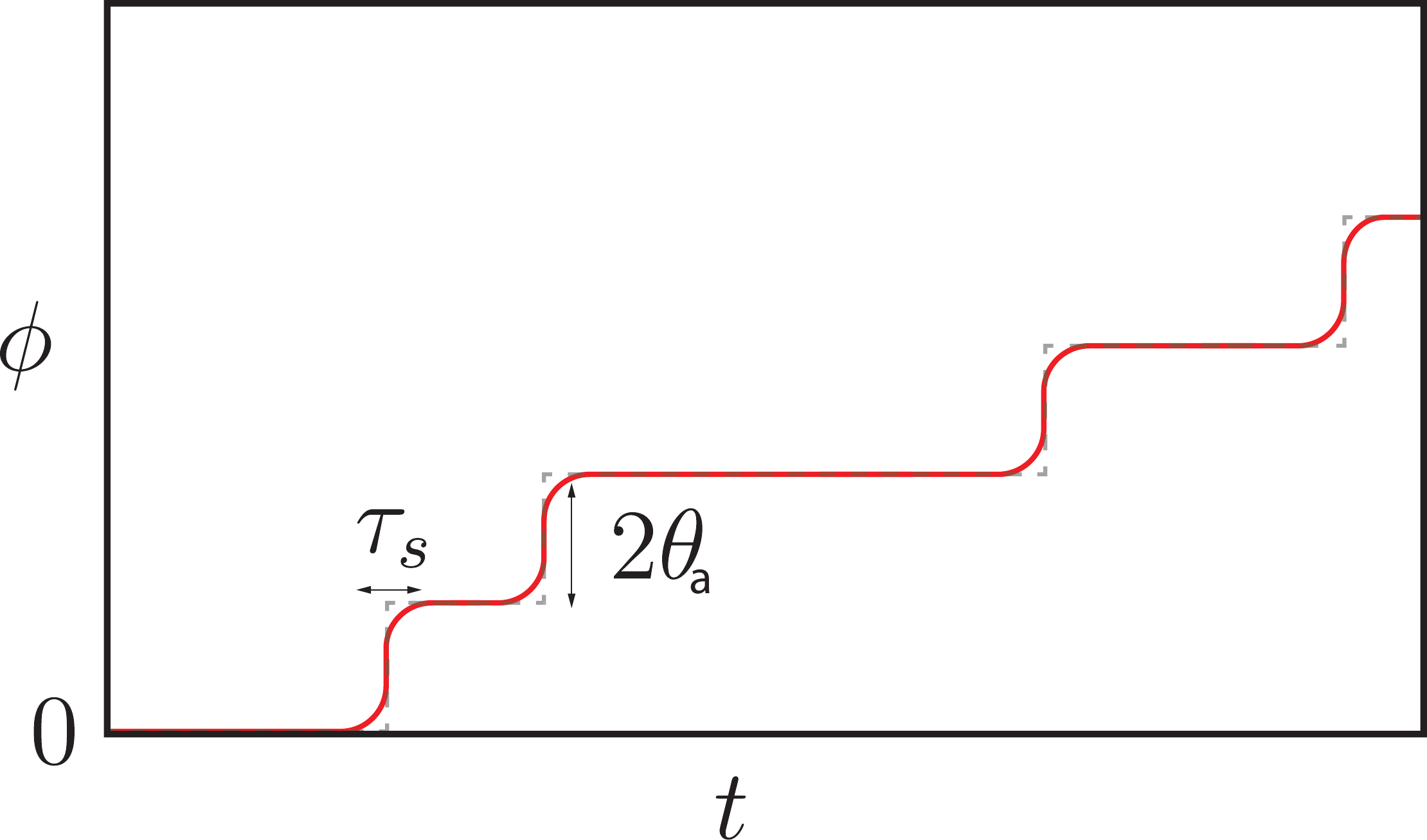}
     \end{minipage}\hfill
      \begin{minipage}[c]{0.48\linewidth}
     \includegraphics[width= \linewidth]{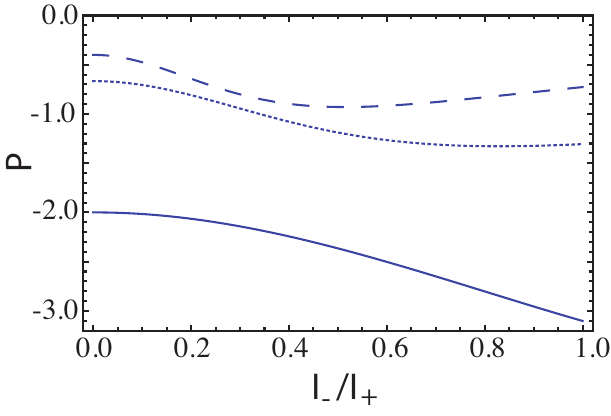}
       \end{minipage}\hfill
    \caption{\label{Fig:BroadenedField}
    Left panel: A step of height $2\theta_a$ in the boson field $\phi$ represents the passage of a quasiparticle with exchange phase $\theta_a$. For $\theta_a>\pi/2$, the finite width $\tau_s$ of the step must be included (adapted from Ref.~\cite{Thamm24}). Right panel: Normalized cross-correlations [Eq.~(\ref{norm_zerobias.eq})] versus relative bias $I_-/I_+$, for $e^*/e = \lambda=\delta=1/m$ and $\theta_a=\pi/m$. Solid line: $m=3$; dotted: $m=5$; dashed: $m=7$ (adapted from Ref.~\cite{Ro+16}). }
\end{figure}

 To connect with experiment, we compute the cross-correlation of the outgoing
currents. Writing $I_u = I_{u,0} - I_T$ and $I_d = I_{d,0} + I_T$, the
cross-correlator takes the form
 %
\begin{eqnarray}
\langle \delta I_d \delta I_u \rangle_{\omega=0} & = & - \langle \delta I_T^2 \rangle_{\omega=0} 
+ e^\star \left( \langle I_{u,0}\rangle {\partial \over 
\partial \langle I_{u,0}\rangle} - \langle I_{d,0}\rangle {\partial \over \partial \langle I_{d,0}\rangle } \right) \langle I_T\rangle  \ ,
\label{FDT.eq}
\end{eqnarray}
%
which is a non-equilibrium fluctuation-dissipation relation: the first term is
the noise generated at the collision QPC, and the second describes the
Johnson-Nyquist-like contribution of incoming current fluctuations transmitted
through the QPC. To better interpret the anyonic contribution we normalize by
the second term on the right-hand side, evaluated at vanishing imbalance,
%
\begin{eqnarray}
P(I_-/I_+) & = & {\langle \delta I_d \delta I_u \rangle_{\omega=0} \over \left. e^\star I_+ {\partial \over \partial I_-} \langle  I_T \rangle \right|_{I_-=0}} \  . \label{noisenormalized.eq}
\end{eqnarray}
%
In the case of two balanced incoming beams ($I_-=0$), this simplifies and may
be written in closed form as  \cite{Ro+16}
%
\begin{equation}
P(0)  =  1 - {\tan \theta_a \over \tan \pi \delta} \, {1 \over 1 - 2 \delta} 
\xrightarrow[\theta_a={\pi \over m}, \delta = {1 \over m}]{} \  {-2 \over m-2} \ \ .
\label{norm_zerobias.eq}
\end{equation}
%

 An important feature of this quantum mechanical result is its universality: the
normalized cross-correlations depend only on the dynamical exponent $\delta$  and the
statistical angle $\theta_a$, but not on microscopic details such as the source
QPC transmissions or the applied voltages separately.  
This robustness makes the prediction well suited for experimental tests. 
The predicted value of the normalized cross-correlation $P(0)$ 
at $\nu=1/3$ has been verified in recent experiments \cite{Bartolomei.2020,Lee.2023,Ruelle.2023,Glidic.2023}.

For the $\nu = 1/3$ Laughlin state, we compare $P(0) = - 2$ with $P_{\rm classical} = - ( 1 - p) T_u$
 (for the case $T_u - T_d$), which would suggest   $p < 0$  describing the bunching of anyons. 
 Although the analogy to a classical exclusion model is instructive, the quantum result is instead governed by time-domain interference of anyonic wave packets, as we will demonstrate in the next section.  The negative correlations therefore 
reflect the coherent braiding phase $\theta_a$, rather than a literal 
two-particle blocking mechanism. The experimental results thereby 
provide direct  evidence for fractional exchange statistics in an 
interferometer-free setting.

  \section{Interpretation in terms of exchange phase and time domain braiding}

We now show that the above results can be interpreted directly in terms of anyonic exchange statistics, following Schiller et al.~\cite{Schiller23}. 
To this end, we use the ${\sf K}$-matrix description for systems with multiple edge modes, which cleanly distinguishes qp charge, filling fraction, and statistical angle.
Concretely, the chiral
boson fields obey
$[\phi_{I}(x), \partial_y \phi_{J}(y ]\ = \  2 \pi i 
\left( K^{-1} \right)_{I,J} \,  
 \delta(x-y) $, 
 where $\phi_I$ are the components  of a vector boson field $\bm{\phi}$, see  \cite{Wen92}.  
The sign of the eigenvalues of the {\bf \sf K}-matrix  determines the direction of propagation of edge modes. 
Distinct topological orders in the quantum Hall regime correspond to inequivalent ${\sf K}$-matrices. 
 In particular, the qp operator associated with an integer vector
$\mathbf l$ is defined as 
$ \psi_{\mathbf l} = e^{i {\mathbf l} ^T  {\bm{ \phi}}}$, 
and the mutual statistical angle between  $\bm{ l}_1$ and  $\bm{l}_2$ quasiparticles is  $\theta_{12} \ = \ \pi \,{\mathbf l}_1^T { \sf K}^{-1} {\mathbf l}_2$. The
electric charge and the filling fraction are given by
$e^\star(\mathbf l)= e\, \mathbf t^T {\sf K}^{-1} \mathbf l$ and
$\nu= \mathbf t^T {\sf K}^{-1} \mathbf t$, respectively, with the charge vector
$\mathbf t$.

From the commutation relations of the boson fields, tunneling of an $\bm{l}_1$ qp at QPC1 generates a kink of magnitude $2\pi K^{-1}\bm{l}_1$ in the boson field $\bm{\phi}$. This induces a phase factor  $e^{2 \pi i {\mathbf l}_2^T K^{-1} {\mathbf l}_1}  \ \equiv \ e^{2 i \theta_{21}}$   in the tunneling operator of an $\bm{l}_2$ quasiparticle at the collision QPC. 
This recovers the factor $e^{2i\theta_a}$ used earlier for Laughlin quasiparticles.
 %
\begin{figure}[t]
  \centering
  \begin{minipage}[t]{0.8\linewidth}
    \includegraphics[width=\linewidth]{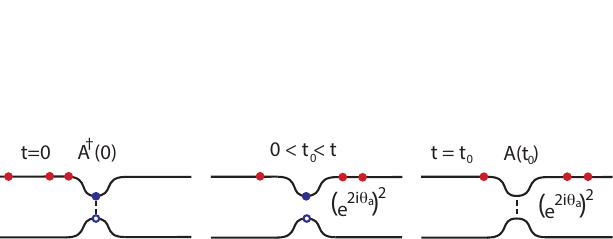}
  \end{minipage}\hfill
  \begin{minipage}[t]{0.15\linewidth}
    \includegraphics[width=\linewidth]{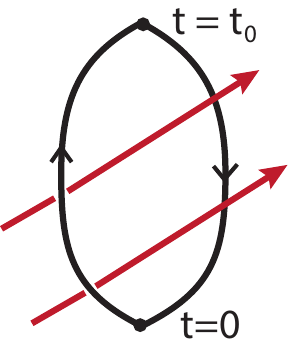}
  \end{minipage}
  \caption{ Time-domain interpretation of the non-equilibrium tunneling correlator $\langle A(t_0) A^\dagger(0)\rangle$. Left panel: At $t=0$, the operator $A^\dagger$ creates a low-energy qp-qh pair at the collision QPC,  propagates until annihilated at $t_0$ by $A$. During the time interval $(0,t_0)$, $N_u = 2$  high-energy anyons arrive from the source QPC, each contributing a phase $e^{2i\theta_a}$.  The power law factor in Eq.~\ref{quantumcorrelation.eq} describes the quantum mechanical amplitude for the qp-qh pair not moving away from the QPC.   Right panel: The qp-qh propagation from $0$ to $t_0$ forms a closed loop in time encircling the arriving anyons, corresponding to a total phase $2\theta_a$ for each anyon which passes the collision QPC.   }
  \label{fig:timedomain}
\end{figure}
%
An alternative and instructive interpretation is provided by the notion of time-domain braiding,  following
Refs.~\cite{Han+16,Lee+20,Lee+22}. To understand this argument, we consider the
correlation function of the tunneling operator
$\langle A(t_0)\, A^\dagger(0)\rangle$, where
$A^\dagger(0)=\zeta^*\,\Psi_u^\dagger(0)\,\Psi_d(0)$ creates, at time $t=0$, a
quasiparticle (qp) on the upper edge and a quasihole (qh) on the lower edge (see Fig.~\ref{fig:timedomain}).
During the interval $(0,t_0)$, $N_u$ qps pass the collision QPC,  before the qp-qh pair is annihilated at $t_0$ by $A(t_0)$. In a
diagrammatic representation, creation and annihilation of the qp-qh pair form a
closed worldline segment in time, which is threaded by the $N_u$ qps that pass
by.  Each threading contributes a phase $e^{2i\theta_a}$, and the accumulated phase $e^{i 2 \theta_a N_u}$ along the closed loop exactly reproduces the non-equilibrium factor   that appears in the generating function of Poissonian
arrivals. 
Thus the process is equivalent to a braiding in the time domain, where a single quasiparticle encircles the impinging quasiparticles in a closed loop;  see
Fig.~\ref{fig:timedomain}. In this picture, the power law factor in Eq.~\ref{quantumcorrelation.eq} describes the quantum mechanical amplitude for the qp-qh pair not moving away from the QPC during the time interval. The same factors appear if one considers the interference contribution between processes where a low energy qp tunnels across the QPC before and after the passage of $N_u$ quasiparticles.
Detailed analysis shows that the process of braiding in time between high-energy injected qps and a low-energy qp drawn from the pre-existing ground state is the dominant process in a collider experiment.
This viewpoint offers a direct and transparent link between the exchange phase $\theta_a$ and the measured cross-correlations, and clarifies why no interferometer is required: the interference arises from time-domain winding of worldlines.

\section{Role of Soliton Broadening}

Collision experiments have recently been extended to hierarchical states, notably the $\nu=2/5$ case involving charge  $e/5$ anyons. 
The results of the experiments by Ruel
et~al.~\cite{Ruelle.2023} and Glidic et~al.~\cite{Glidic.2023} are consistent
with each other and report a negative Fano factor, with a magnitude smaller than that observed at $\nu=1/3$.
A standard theoretical interpretation models the $\nu=2/5$ state as  a hierarchical FQH state \cite{Halperin84,Haldane83,Jain89}: The
$\nu=2/5$ state is modeled as a maximum-density condensate of charge-$1/3$
quasiparticles on top of a $\nu=1/3$ Laughlin state, with a total density
corresponding to $\nu=2/5$. 
With appropriate confinement, the edge structure can support
a $1/3$ edge mode between vacuum and the
$1/3$ parent state, and another edge mode between fillings $1/3$ and $2/5$,
which are independent from each other when their distance is much larger than the
magnetic length. Each of these modes can be described using the hydrodynamic
approach \cite{Wen92}. In this model, there are independent $1/3$ and $1/15$
edge channels, giving rise to an anyonic exchange phase $\theta_a=3\pi/5$
between $e^\ast=1/5$ quasiparticles, and a dynamical exponent $\delta=3/5$ for tunneling of $1/5$ quasiparticles between $1/15$ edge channels. 
Direct application of Eq.~(\ref{norm_zerobias.eq}) incorrectly predicts a large positive Fano factor, contradicting experiment.

Refs.~\cite{Schiller23,Thamm24,Yer+24} emphasized that the finite temporal width of anyonic solitons is essential for correctly describing exchange phases $\theta_a > \pi/2$. Treating quasiparticles as infinitely sharp steps introduces ambiguities in the statistical phase and yields incorrect Fano factors.
Including a finite soliton width resolves these 
ambiguities and restores agreement with experiment. 
The discrepancy arises because the step-function soliton model fails for large exchange phases, 
owing to an ambiguity between assigning a phase factor $e^{2i\theta_a}$ (upward step) or $e^{-2i(\pi-\theta_a)}$ (downward step).

The resolution is to broaden the non-equilibrium field steps, ensuring monotonic and unambiguous phase accumulation;  see
Refs.~\cite{Schiller23,Thamm24}. In this framework, a soliton of width $\tau_s$
can be described by an arctangent function
\begin{align}
   \phi_{{\rm noneq},\alpha}(t) =  2 \theta_a \left[   \frac{1}{\pi}\arctan\left(\frac{t-t_0}{\tau_s}\right) + \frac{1}{2} \right] \ , \label{Eq:BosFieldSingPart}
\end{align}
thereby regularizing the short-time correlations and eliminating the sign ambiguity  for $\theta_a>\pi/2$.

The Fano factor is obtained as a double integral over this soliton shape
function and can be evaluated using the framework of Ref.~\cite{Thamm24} (for
related work see Ref.~\cite{Yer+24}).  The generalized Fano factor depends not only on $\theta_a$, but also on the dynamical exponent $\delta$ and the dilution $T_s \equiv \frac{I_+}{e^\ast}\tau_s$, which is approximately equal to the
transparency of the source QPC \cite{Thamm24}.
For $\delta>1/2$, short-time fluctuations are significant, and the soliton width $\tau_s$ explicitly enters the result, 
 explaining why experiments on $\nu=2/5$ states produce weakly negative Fano factors, contrasting with the strong positive values predicted by the step-function model.

This formalism now allows one to describe charge-$e/5$ anyons and 
gives $P(0)=-0.21$ for balanced beams,  a dynamical exponent
$\delta=3/5$,  and a dilution $T_s=0.1$ \cite{Thamm24}.  Experiments \cite{Ruelle.2023,Glidic.2023} find a somewhat larger negative value, but overall the agreement is reasonable.

\section{Conclusions}

We have presented the anyon collider as a revealing probe of fractional statistics. 
In this geometry, dilute quasiparticle beams
produced at two source QPCs collide at a third QPC, and the cross-correlations
of the outgoing currents depend on the statistical angle. This experiment can be analyzed using a non-equilibrium bosonization framework.   For the Laughlin states at $\nu=1/m$, the normalized cross-correlations are universal: they depend only on the current imbalance, the dynamical exponent, and the statistical angle, but not on the transmission probability of  source QPCs.
This universality enables measurement of the statistical angle $\theta_a$, although quantitative determination requires knowledge of the dynamical exponent $\delta$.  

A central point is that the signal arises from time-domain interference between the injected high-energy anyons and low-energy anyons, which tunnel back and forth at the QPC and are present, virtually, in the system's ground state.  The correlations reflect braiding phases accumulated by worldlines winding in time, not from spatial exclusion or conventional interferometry.  This viewpoint is compatible with a ${\sf K}$-matrix description
of multi-mode edges, where an injected quasiparticle modifies the tunneling operator at the collision point by a phase set by the mutual statistics.

Extension to hierarchical states reveals the crucial role of finite pulse shapes. 
For $\theta_a>\pi/2$, step-function models of quasiparticles produce phase ambiguities and erroneous Fano factors.
Including a finite soliton width resolves this problem and regularizes the short-time physics. In particular, it provides a quantitative description of charge-$e/5$ quasiparticle collisions at $\nu=2/5$ and explains the weakly negative Fano factors observed experimentally, highlighting their dependence on the dynamical exponent and beam dilution.

\address{Institut f\"ur Theoretische Physik, Universit\"at  Leipzig\\
04103  Leipzig, Germany.\\
\email{rosenow@physik.uni-leipzig.de}}

\address{Department of Physics, Harvard University\\
 Cambridge, MA 02138, USA.\\
\email{halperin@g.harvard.edu}}

\end{document}